\documentstyle[aps]{revtex}

\begin{document}
\draft
\title{Quantum effects in the collective light scattering by coherent atomic recoil in a 
Bose-Einstein condensate}
\author{N.\ Piovella, M.\ Gatelli and R.\ Bonifacio\\
\it 
Dipartimento di Fisica, Universit\`a Degli Studi di Milano, INFN \&
INFM\\ Via Celoria 16, Milano I-20133,Italy}

\maketitle

\begin{abstract}
We extend the semiclassical model of the collective atomic recoil laser (CARL) to include
the quantum mechanical description of the center-of-mass motion of the atoms in a 
Bose-Einstein condensate (BEC). We show that when the average atomic momentum 
is less than the recoil momentum $\hbar\vec q$, the CARL equations reduce to the Maxwell-Bloch 
equations for two momentum levels. In the conservative regime (no radiation losses), the 
quantum model depends on a single collective parameter, $\rho$, that can be 
interpreted as the average number of photons scattered per atom in the classical limit.
When $\rho\gg 1$, the semiclassical CARL regime is recovered, with many momentum levels 
populated at saturation. On the contrary, when $\rho\le 1$, the average momentum oscillates 
between zero and $\hbar\vec q$, and a periodic train of $2\pi$ hyperbolic secant pulses is 
emitted. 
In the dissipative regime (large radiation losses) and in a suitable quantum limit, a 
sequential superfluorescence scattering occurs, in which after each process atoms emit a 
$\pi$ hyperbolic secant pulse and populate a lower momentum state. 
These results describe the regular arrangement of the momentum pattern observed in recent 
experiments of superradiant Rayleigh scattering from a BEC.
\end{abstract}

With the realization of Bose Einstein condensation (BEC) in dilute alkali gases
\cite{BEC}, it is now possible to study the coherent interaction between light and an
ensemble of atoms prepared in a single quantum state. For example, Bragg diffraction
\cite{Kozuma} of a Bose-Einstein condensate  by a moving optical standing wave can be 
used to diffract any fraction of a BEC into a selectable momentum state, realizing an atomic 
beam splitter. Among the multitude of experiments studying the behavior of a BEC under the 
action of external laser beams, only a small number have been devoted to the 
active role caused by the atoms in the condensate on the radiation.
In particular, collective light scattering and matter-wave amplification 
caused by coherent center-of mass motion of atoms in a condensate illuminated by a far 
off-resonant laser were recently observed\cite{MIT:1,MIT:2,Tokio}. 
These experiments have been interpreted in Ref.\cite{MIT:1} as 
superradiant Rayleigh scattering, and successively investigated in Ref.
\cite{Meystre:1} and \cite{Ozgur} using a quantum theory based on a quantum 
multi-mode extention of the collective atomic recoil laser (CARL) Hamiltonian 
model originally derived by Bonifacio {et al.}\cite{CARL:1,CARL:2,CARL:3,CARL:4}. 
In particular, the original semiclassical CARL model was extended in 
Ref.\cite{Meystre:2,Meystre:3} to include a quantum mechanical description of the 
center-of-mass motion of the atoms in the condensate.
Whereas the analysis of ref.\cite{Meystre:2,Meystre:3} is limited to the study of the 
onset of the collective instability starting from quantum fluctuations, 
some nonlinear effects due to momentum population depletion were discussed in 
ref.\cite{Meystre:1} and \cite{Ozgur}.

Recently, we have shown\cite{OC} that the superradiant Rayleigh scattering from a BEC can be 
satisfactorily interpreted in terms of the CARL mechanism using a semiclassical model in the 
'mean-field' approximation\cite{SF}, in which the rapid escape of the radiation from the 
condensate is modelled by a decay of the field amplitude at the rate $\kappa_c=c/2L$, where 
$L$ is the sample length. The main drawback of the semiclassical model is that, as it 
considers the center-of-mass motion of the atoms as classical, it cannot describe the 
discreteness of the recoil velocity, as has been observed in the experiment of 
Ref.\cite{MIT:1}.

The aim of this work is to extend the semiclassical CARL model to include the 
quantum mechanical description of the center-of-mass motion of a sample of cold atoms.
The quantum model that we obtain is equivalent to that derived by Moore and Meystre\cite{Meystre:2} 
using second quantization techniques. However, whereas the work of Ref.\cite{Meystre:2} is 
focussed on the linear regime and on the start-up of the instability, we study the full 
nonlinear regime and the quantum and classical limits of the model. Our basic result is that 
the atomic motion is quantized when the average recoil momentum is comparable to 
$\hbar\vec q$ (where $\vec q=\vec k_2-\vec k_1$ is 
the difference between the incident and the scattered wave vectors), i.e. the recoil momentum
gained by the atom trading a photon via absorbtion and stimulated emission between the incident 
and scattered waves.
In this limit, the quantum CARL equations reduce to the Maxwell-Bloch equations for 
two momentum levels\cite{MIT:3}. In the 'conservative' (or 'hamiltonian') regime, 
in which the radiation losses are negligible, this occurs for $\rho<1$, where the CARL 
parameter $\rho$ represents the average number of photons 
scattered per atom in the classical limit. In the superradiant regime, for $\kappa>1$ 
(where $\kappa=\kappa_c/\omega_r\rho$, $\kappa_c$ is the radiation loss, $\omega_r\rho$ 
is the collective recoil bandwidth, $\omega_r=\hbar |\vec q|^2/2M$ is the recoil frequency and 
$M$ is the atomic mass), the atomic motion becomes quantized for $\rho<\sqrt{2\kappa}$.
In this limit, we demonstrate that a sequential superfluorescence (SF) scattering occurs, 
in which, during each process, the atoms emit a $\pi$ hyperbolic secant 
pulse and populate a lower momentum level, as it has been observed in the MIT 
experiment\cite{MIT:1}.

Our starting point is the classical model of equations for $N$ two-level atoms exposed to an 
off-resonant pump laser, whose electric field 
$\vec E_0=\hat e{\cal E}_0\cos(\vec k_2\cdot\vec x-\omega_2t)$ is polarized along $\hat e$,
propagates along the direction of $\vec k_2$ and has a frequency 
$\omega_2=ck_2$ with a detuning from the atomic resonance, $\Delta_{20}=\omega_2-\omega_0$, 
much larger than the natural linewidth of the atomic transition, $\gamma$.
We assume the presence of a scattered field ('probe beam') with frequency 
$\omega_1=\omega_2-\Delta_{21}$, wavenumber $\vec k_1$ making an angle $\phi$ with 
$\vec k_2$ and electric field
$\vec E=(\hat e/2)[{\cal E}(t)e^{i(\vec k_1\cdot\vec x-\omega_1 t)}+{\rm c.c.}]$
with the same polarization of the pump field.
In the absence of an injected probe field, the emission starts from fluctuations and the 
propagation direction of the scattered field is determined either by the geometry of the 
condensate (as in the case of the MIT experiment\cite{MIT:1}, where the condensate has a cigar 
shape) or by the presence of an optical resonator tuned on a selected longitudinal mode. 
By adiabatically eliminating the internal atomic degree of freedom, the following semiclassical 
CARL equations has been derived\cite{CARL:1,CARL:2,CARL:3}:
\begin{eqnarray}
\frac{d\theta_j}{d\tau}&=&\overline p_j\label{CARL1}\\
\frac{d\overline p_j}{d\tau}&=&-\left[\tilde Ae^{i\theta_j}+{\rm c.c.}\right]\label{CARL2}\\
\frac{d\tilde A}{d\tau}&=&\frac{1}{N}\sum_{j=1}^N e^{-i\theta_j}+i\delta\tilde A\label{CARL3}
\end{eqnarray}
where $\tau=\rho\omega_r t$ is the interaction time in units of $\omega_r\rho$, 
$\theta_j=(\vec k_1-\vec k_2)\cdot\vec x_j=qz_j$ and $\overline p_j=qv_{zj}/\rho\omega_r$ 
(where $q=|\vec q|$) are the dimensionless position and velocity of the $j$-th atom along the 
axis $\hat z$,
$\tilde A=-i(\epsilon_0/n_s\hbar\omega\rho)^{1/2}{\cal E}(\tau)e^{i\delta\tau}$,
$\delta=\Delta_{21}/\omega_r\rho$ and
$\rho=(\Omega_0/2\Delta_{20})^{2/3}(\omega\mu^2 n_s/\hbar\epsilon_0\omega_r^2)^{1/3}$ 
is the collective CARL parameter. 
$\Omega_0=\mu{\cal E}_0/\hbar$ is the Rabi frequency of the pump,
$n_s=N/V$ is the average atomic density of the sample (containing $N$ atoms in a volume 
$V$) and $\mu$ is the dipole matrix element.
We assume $\omega_2\approx\omega_1=ck$, so that $q\approx 2k\sin^2(\phi/2)$.
Eqs.(\ref{CARL1})-(\ref{CARL3}) are formally equivalent to those of the free electron laser 
model\cite{FEL}.
In order to quantize both the radiation field and the center-of-mass motion of the atoms,
we consider $\theta_j$, $p_j=(\rho/2){\overline p}_j=Mv_{zj}/\hbar q$
and $a=(N\rho/2)^{1/2}\tilde A$ as quantum operators satisfying the canonical commutation 
relations $[\theta_j,p_{j'}]=i\delta_{jj'}$ and $[a,a^{\dag}]=1$. 
With these definitions, Eqs.(\ref{CARL1})-(\ref{CARL3}) are the Heisenberg equations of motion
associated with the Hamiltonian:
\begin{equation}
H=\frac{1}{\rho}
\sum_{j=1}^N p_j^2+ig
\left(\sum_{j=1}^N a^{\dag}e^{-i\theta_j}-{\rm h.c.}\right)
-\delta a^{\dag}\tilde a=\sum_{j=1}^N H_j(\theta_j,p_j),
\label{ham}
\end{equation}
where $g=\sqrt{\rho/2N}$. We note that $[H,Q]=0$, where $Q=a^{\dag}a+\sum_{j=1}^Np_j$ is the 
total momentum in units of $\hbar q$.
In order to obtain a simplified description of a BEC as a system of $N$ noninteracting atoms
in the ground state, we use the Schr\"odinger picture for the atoms 
(instead of the usual Heisenberg picture\cite{RB}), i.e. 
$|\psi(\theta_1,\dots,\theta_N)\rangle=|\psi(\theta_1)\rangle\ldots|\psi(\theta_N)\rangle$, 
where $|\psi(\theta_j)\rangle$ obeys the single-particle Schr\"odinger equation,
$i(\partial/\partial\tau)|\psi(\theta_j)\rangle=H_j(\theta_j,p_j)|\psi(\theta_j)\rangle$.
In this paper we describe the light field classically. Hence, considering the field operator 
$a$ as a c-number, eq.(\ref{CARL3}) yields:
\begin{equation}
\frac{da}{d\tau}=i\delta a+g\sum_{j=1}^N
\langle\psi(\theta_j)|e^{-i\theta_j}|\psi(\theta_j)\rangle.
\label{a1}
\end{equation}
Let now expand the single-atom wavefunction on the momentum basis, 
$|\psi(\theta_j)\rangle=\sum_n c_j(n)|n\rangle_j$, where $p_j|n\rangle_j=n|n\rangle_j$, 
$n=-\infty,\dots\infty$ and $c_j(n)$ is the probability amplitude of the $j$-th atom having
momentum $-n\hbar\vec q$. Introducing the collective density matrix:
\begin{equation}
S_{m,n}=\frac{1}{N}\sum_{j=1}^N c_j(m)^*c_j(n)e^{i(m-n)\delta\tau},
\label{matrix}
\end{equation}
a straightforward calculation yields, from Eqs.(\ref{ham}) and (\ref{a1}),
the following closed set of equations:
\begin{eqnarray}
\frac{dS_{m,n}}{d\tau}&=&
i(m-n)\delta_{m,n}S_{m,n}+\frac{\rho}{2}
\left[A\left(S_{m+1,n}-S_{m,n-1}\right)+
A^*\left(S_{m,n+1}-S_{m-1,n}\right)\right]\label{s3}\\
\frac{dA}{d\tau}&=&\sum_{n=-\infty}^{\infty}S_{n,n+1}-\kappa A
\label{a3},
\end{eqnarray}
where $\delta_{m,n}=\delta+(m+n)/\rho$ and we have redefined the field as 
$A=\sqrt{2/\rho N}a e^{-i\delta\tau}$. We have also introduced a
damping term $-\kappa A$ in the field equation, where $\kappa=\kappa_c/\omega_r\rho$, 
$\kappa_c=c/2L$ and $L$ is the sample length along the probe propagation, which provides an
approximated model describing the escape of photons from the atomic medium.
In the presence of a ring cavity of length $L_{\rm cav}$ and reflectivity $R$, 
$\kappa_c=-(c/L_{\rm cav}){\rm ln}R$, as shown in the usual 'mean-field' 
approximation\cite{CARL:4}.
Eqs.(\ref{s3}) and (\ref{a3}) are completely equivalent to the CARL equations 
(\ref{CARL1})-(\ref{CARL3}) and determine the temporal evolution of the density matrix elements
for the momentum levels. In particular, $p_n=S_{n,n}$ is the probability of finding the atom 
in momentum level $|n\rangle$, $\langle p\rangle=\sum_n nS_{n,n}$ is the average momentum and 
$\sum_n S_{n,n+1}$ is the bunching parameter. Eqs.(\ref{s3}) and (\ref{a3}) are identical to 
these derived by Moore and coworkers\cite{Meystre:2} second quantizing the single-particle 
Hamiltonian $H_j$ and introducing bosonic creation and annihilation operators of a given 
center-of-mass momentum. 
For a constant field $A$, Eq.(\ref{s3}) describes a Bragg scattering process, in which 
$m-n$ photons are absorbed from the pump and scattered into the probe, changing the initial 
and final momentum states of the atom from $m$ to $n$. 
Conservation of energy and momentum require that during this process
$\omega_1-\omega_2=(m+n)\omega_r$, i.e. $\delta_{m,n}=0$.
Eqs.(\ref{s3}) and (\ref{a3}) conserve the norm, i.e. $\sum_m S_{m,m}=1$, and,
when $\kappa=0$, also the total momentum $Q=(\rho/2)|A|^2+\langle p\rangle$.
Figure \ref{fig1}a shows $|A|^2$ vs. $\tau$, for $\kappa=0$, $\delta=0$ and $A(0)=10^{-4}$,
comparing the semiclassical solution with the quantum solution in the classical limit, 
$\rho\gg 1$: the dashed line is the numerical solution of Eqs.(\ref{CARL1})-(\ref{CARL3}),
for a classical system of $N=200$ cold atoms, with initial momentum $p_j(0)=0$ 
(where $j=1,\dots,N$) and phase $\theta_j(0)$ uniformly distributed over $2\pi$, i.e. unbunched;
the continuous line is the numerical solution of Eqs.(\ref{s3}) and (\ref{a3}) for $\rho=10$ 
and a quantum system of atoms initially in the ground state $n=0$, i.e. with 
$S_{n,m}=\delta_{n0}\delta_{m0}$. 
Figure \ref{fig1}a shows that the quantum system behaves, with good approximation, 
classically. Because from Fig.\ref{fig1}a the maximum dimensionless intensity is 
$|A|^2\approx 1.4$, the constant of motion $Q$ gives $\langle p\rangle\approx -0.7\rho$
and the maximum average number of emitted photons is about $\langle a^{\dag}a\rangle\sim N\rho$. 
Hence, the CARL parameter $\rho$ can be interpreted as the maximum average number of photons 
emitted per atom (or equivalently, as the maximum average momentum 
recoil, in units of $\hbar q$, acquired by the atom) in the classical limit. 
Figure \ref{fig1}b shows the distribution of the population level $p_n$ at the first peak of 
the intensity of fig.\ref{fig1}a, for $\tau=12.4$. 
We observe that, at saturation, twenty-five momentum levels are occupied, with an induced
momentum spread comparable to the average momentum.

Let us now consider the equilibrium state with no probe field, $A=0$, and all the atoms in the 
same momentum state $n$, i.e. with $S_{n,n}=1$ and the other matrix elements zero. This is 
equivalent to assume the temperature of the system equal to zero and all the atoms moving with 
the same velocity $-n\hbar\vec q$, without spread. 
This equilibrium state is unstable for certain values of the
detuning. In fact, by linearizing Eqs.(\ref{s3}) and (\ref{a3})
around the equilibrium state, the only matrix elements giving linear
contributions are $S_{n-1,n}$ and $S_{n,n+1}$, showing that in the linear regime
the only transitions allowed from the state $n$ are these towards the levels $n-1$ and
$n+1$. Introducing the new variables $B_n=S_{n,n+1}+S_{n-1,n}$ and 
$P_n=S_{n,n+1}-S_{n-1,n}$, Eqs.(\ref{s3}) and (\ref{a3}) reduce to the linearized equations:
\begin{eqnarray}
\frac{dB_n}{d\tau}&=&-i\delta_nB_n-\frac{i}{\rho}P_n\label{l1}\\
\frac{dP_n}{d\tau}&=&-i\delta_nP_n-\frac{i}{\rho}B_n-\rho A
\label{l2}\\
\frac{d A}{d\tau}&=&B_n-\kappa A
\label{l3},
\end{eqnarray}
where $\delta_n=\delta+2n/\rho$. Seeking solutions proportional to 
$e^{i(\lambda-\delta_n)\tau}$, we obtain the following cubic dispersion relation:
\begin{equation}
\left(\lambda-\delta_n-i\kappa\right)
\left(\lambda^2-1/\rho^2\right)+1=0.
\label{cubic}
\end{equation}
In the exponential regime, when the unstable (complex) root $\lambda$ dominates, 
$B(\tau)\sim e^{i(\lambda-\delta_n)\tau}$ and, from Eq.(\ref{l1}), $P_n=-\rho\lambda B_n$.
The semiclassical limit is recovered for $\rho\gg 1$ (when $\kappa=0$) or
$\rho\gg\sqrt{\kappa}$ (when $\kappa>1$) and 
$\delta_n\approx\delta$, i.e. neglecting the shift due to the recoil frequency $\omega_r$.
In this limit, maximum gain occurs for $\delta=0$, with $\lambda=(1-i\sqrt 3)/2$ when 
$\kappa=0$ or $\lambda=-(1+i)/\sqrt{2\kappa}$ when $\kappa>1$. Furthermore, 
$|S_{n,n+1}|\sim|S_{n-1,n}|$, so that the atoms may experience both emission and absorbtion.
This result can be interpreted in terms of single-photon emission and absorpion by an atom 
with initial momentum $-n\hbar\vec q$. In fact, energy and momentum conservation impose 
$\omega_1-\omega_2=(2n\mp 1)\omega_r$ (i.e. $\delta_n=\pm 1/\rho$) when a probe photon
is emitted or absorbed, respectively. Because in the semiclassical limit the gain 
bandwidth is $\Delta\omega\sim \omega_r\rho\gg \omega_r$ when $\kappa=0$ (or
$\Delta\omega\sim\kappa_c\gg \omega_r$ when $\kappa>1$) the atom can both emit or 
absorbe a probe photon.
On the contrary, in the quantum limit the recoil energy $\hbar\omega_r$ can not be neglected,
and there is emission without absorbtion if $|S_{n,n+1}|\ll |S_{n-1,n}|$, i.e. 
$B_n\approx -P_n$ and $\lambda\approx 1/\rho$.
This is true for $\rho<1$ when $\kappa=0$, with the unstable root 
$\lambda\approx 1/\rho+\delta_n'/2-1/2\sqrt{\delta_n'^2-2\rho}$ (where
$\delta_n'=\delta_n-1/\rho$), and for $\rho<\sqrt{2\kappa}$ when $\kappa>1$, 
with ${\rm Re}\lambda\approx 1/\rho+(\rho\delta_n'/2)/(\delta_n'^2+\kappa^2)$ and
${\rm Im}\lambda\approx -(\rho\kappa/2)/(\delta_n'^2+\kappa^2)$. In both cases,
maximum gain occurs for $\delta_n=1/\rho$ (i.e. $\Delta_{21}=(1-2n)\omega_r$) within a
bandwidth $\Delta\omega\sim\omega_r\rho^{3/2}$ and $\Delta\omega\sim\omega_r\rho^{2}/\kappa$
(respectively for $\kappa=0$ and $\kappa>1$), which are both less than the frequency difference
$2\omega_r$ between the emission and absorbtion lines. Hence, in the quantum limit the optical
gain is due exclusively to emission of photons, whereas in the semiclassical limit gain
results from a positive difference between the average emission and absorbtion rates.
When $\kappa=0$, the resonant gain in the limit $\rho< 1$ is
$G_S=\omega_r\rho\sqrt{\rho/2}=\sqrt{3/8\pi}(\Omega_0/2\Delta_{20})\gamma\sqrt{N_{\rm eff}}$, 
where $\gamma=\mu^2k^3/3\pi\hbar\epsilon_0$ is the natural decay rate of the atomic transition
and $N_{\rm eff}=(\lambda^2/A)(c/\gamma L)N$ is the effective atomic number in the volume 
$V=AL$, where $A$ and $L$ are the cross section and the length of the sample. When $\kappa>1$, 
the resonant SF gain in the limit $\rho< \sqrt{2\kappa}$ is
$G_{\rm SF}=\omega_r\rho^2/2\kappa=(3/4\pi)\gamma(\Omega_0/2\Delta_{20})^2(\lambda^2/A)N$.

The above results show that the combined effect of the probe and pump fields on a collection
of cold atoms in a pure momentum state $n$ is responsible of a collective instability that leads
the atoms to populate the adjacent momentum levels $n-1$ and $n+1$. 
However, in the quantum limit $\rho<1$ when $\kappa=0$ 
(or $\rho<\sqrt{2\kappa}$ when $\kappa>1$) conservation of energy and momentum of the photon
constrains the atoms to populate only the lower momentum level $n-1$. 
This holds also in the nonlinear regime, as we have verified solving numerically 
Eqs.(\ref{s3}) and (\ref{a3}).
In the quantum limit above, the exact equations reduce to those for only three matrix 
elements, $S_{n,n}$, $S_{n-1,n-1}$ and $S_{n-1,n}$, with $S_{n-1,n-1}+S_{n,n}=1$.
Introducing the new variables $S_n=S_{n-1,n}$ and $W_n=S_{n,n}-S_{n-1,n-1}$, 
Eqs.(\ref{s3}) and (\ref{a3}) reduce to the well-known Maxwell-Bloch equations\cite{MBE}:
\begin{eqnarray}
\frac{dS_n}{d\tau}&=&-i\delta_n'S_n+\frac{\rho}{2}AW_n  	\label{MB1}\\
\frac{dW_n}{d\tau}&=&-\rho(A^* S_n+{\rm h.c.})       	\label{MB2}\\
\frac{d A}{d\tau}&=&S_n-\kappa A					\label{MB3},
\end{eqnarray}
where $\delta_n'=\delta+(2n-1)/\rho$. 
When $\kappa=0$ and $\delta_n'=0$, if the system starts radiating incoherently by pure 
quantum-mechanical spontaneous emission, the solution of Eqs.(\ref{MB1})-(\ref{MB3}) is 
a periodic train of $2\pi$ hyperbolic secant pulses\cite{BP} with
$|A|^2=(2/\rho){\rm sech}^2[\sqrt{\rho/2}(\tau-\tau_n)]$, 
where $\tau_n=(2n+1){\rm ln}(\rho/2)/\sqrt{\rho/2}$. Furthermore, 
the average momentum $\langle p\rangle=n+{\rm Th}^2[\sqrt{\rho/2}(\tau-\tau_n)]-1$
oscillates between $n$ and $n-1$ with period $\tau_n$. We observe that the maximum number
of photons emitted is $\langle a^{\dag}a\rangle_{\rm peak}=(\rho N/2)|A|^2_{\rm peak}=N$, as 
expected. Figure \ref{fig2} shows the results of a numerical integration of 
Eqs.(\ref{s3}) and (\ref{a3}), for $\kappa=0$, $\rho=0.2$ and $\delta=5$, with the atoms 
initially in the momentum level $n=0$ and the field starting from the seed value $A_0=10^{-5}$.
Figures \ref{fig2}a and b show the
intensity $|A|^2$ and the average momentum $\langle p\rangle$ vs. $\tau$,
in agreement with the predictions of the reduced Eqs.(\ref{MB1})-(\ref{MB3}).

In the superradiant regime, $\kappa>1$, Eqs.(\ref{MB1})-(\ref{MB3}) describe a single 
SF scattering process in which the atoms, initially in the momentum state $n$, 'decay' to the 
lower level $n-1$ emitting a $\pi$ hyperbolic secant pulse, with intensity
$|A|^2=1/[4(\kappa^2+\delta_n'^2)]
{\rm sech}^2[(\tau-\tau_D)/\tau_{SF}]$ and average momentum
$\langle p\rangle=n-(1/2)\{1+{\rm Th}[(\tau-\tau_D)/\tau_{SF}]\}$,
where $\tau_{SF}=2(\kappa^2+\delta_n'^2)/\rho\kappa$ is the 'superfluorescence time'\cite{SF},
$\tau_D=\tau_{SF}{\rm Arcsech}(2|S_n(0)|)\approx-\tau_{SF}{\rm ln}\sqrt{2|S_n(0)|}$ is the 
delay time and $|S_n(0)|\ll 1$ is the initial polarization.
Figures \ref{fig3}a and b shows $|A|^2$ and $\langle p\rangle$ vs. $\tau$ calculated solving
Eqs.(\ref{s3}) and (\ref{a3}) numerically with $\kappa=10$, $\rho=2$, $\delta=0.5$ 
and the same initial conditions of fig. \ref{fig2}. We observe a sequential SF
scattering, in which the atoms, initially in the level $n=0$, change their momentum by discrete
steps of $\hbar\vec q$ and emit a SF pulse during each scattering process. We observe that 
for $\delta=1/\rho$ the field is resonant only with the first transition, 
from $n=0$ to $n=-1$; for a generic initial state $n$, resonance occurs when
$\delta=(1-2n)/\rho$, so that in the case of figure \ref{fig3}a the peak intensity of the 
successive SF pulses is reduced (by the factor $1/[\kappa^2+(2n/\rho)^2]$)
whereas the duration and the delay of the pulse are increased. However, the pulse retains
the characteristic sech$^2$ shape and the area remains equal to $\pi$, inducing the atoms
to decrease their momentum by a finite value $\hbar\vec q$.
We note that, although the SF time in the quantum limit ($\tau_{SF}=2\kappa/\rho$ at resonance)
can be considerable longer than the characteristic superradiant time obtained in the 
classical limit, $\tau_{SR}=\sqrt{2\kappa}$, the peak intensity of the pulse in the
quantum limit is always approximately half of the value obtained in the semiclassical limit
(see Ref.\cite{OC} for details).

In conclusion, we have shown that the CARL model describing a system of atoms in their momentum 
ground state (as those obtained in a BEC) and properly extended to include a quantum-mechanical 
description of the center-of-mass motion, allows for a quantum limit in which the average 
atomic momentum changes in discrete units of the photon recoil momentum $\hbar\vec q$ and reduce
to the Maxwell-Bloch equations for two momentum levels. These results demonstrate that the
regular arrangement of momentum pattern observed  in the MIT experiment\cite{MIT:1} can be 
interpreted as being due to sequential superfluorescence scattering. A detailed study of this 
and other aspects of the MIT experiment will be the object of a future extended pubblication.

We thank G.R.M. Robb for a careful reading of the manuscript and for helpful discussions.

\begin{figure}
\caption{Classical limit of CARL for $\rho\gg 1$ in the case $\kappa=0$.
(a): $|A|^2$ vs. $\tau$ as obtained from the classical eqs.(1)-(3) (dashed line) and from the
quantum eqs.(7) and (8) for $\rho=10$ (solid line);
(b): population level $p_n$ vs. $n$ at the occurring of the first maximum of $|A|^2$, 
at $\tau=12.4$. The other parameters are $\delta=0$ and $A(0)=10^{-4}$.}
\label{fig1}
\end{figure}

\begin{figure}
\caption{Quantum limit of CARL for $\rho< 1$ in the case $\kappa=0$.
(a) $|A|^2$ and (b) $\langle p\rangle$ vs. $\tau$, for $\rho=0.2$, $\delta=5$,
$A(0)=10^{-5}$ and the atoms initially in the state $n=0$.
We note that $\langle p\rangle=-(2/\rho)(|A|^2-|A(0)|^2)$.}
\label{fig2}
\end{figure}

\begin{figure}
\caption{Sequential superfluorescent (SF) regime of CARL.
(a) $|A|^2$ and (b) $\langle p\rangle$ vs. $\tau$, for $\rho=2$, $\delta=0.5$, 
$\kappa=10$, and the same initial conditions of fig.2.}
\label{fig3}
\end{figure}

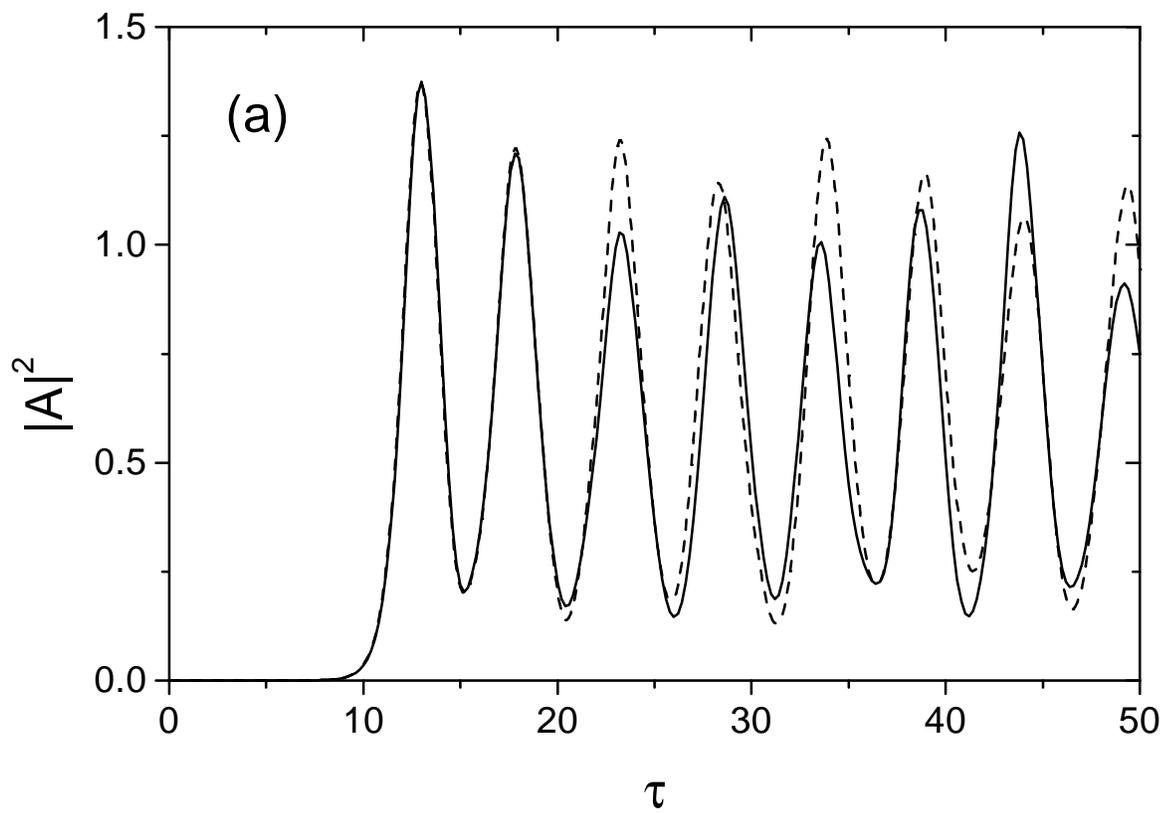
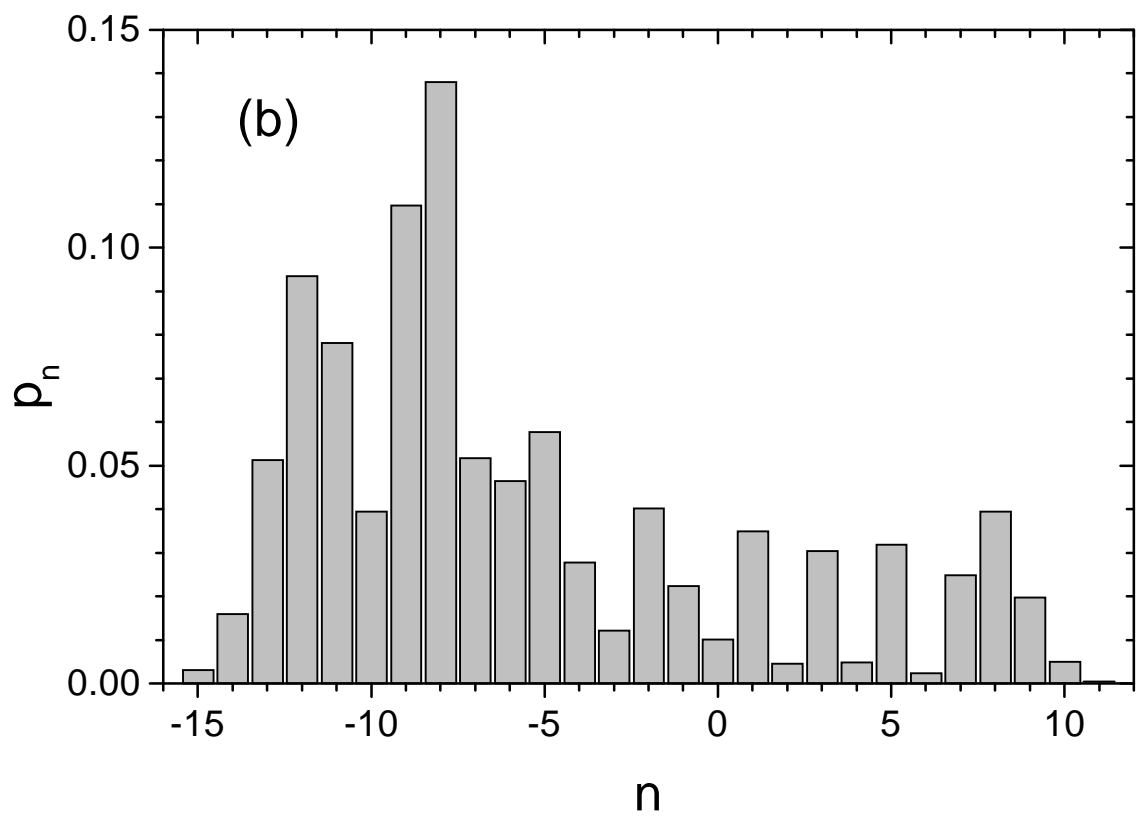
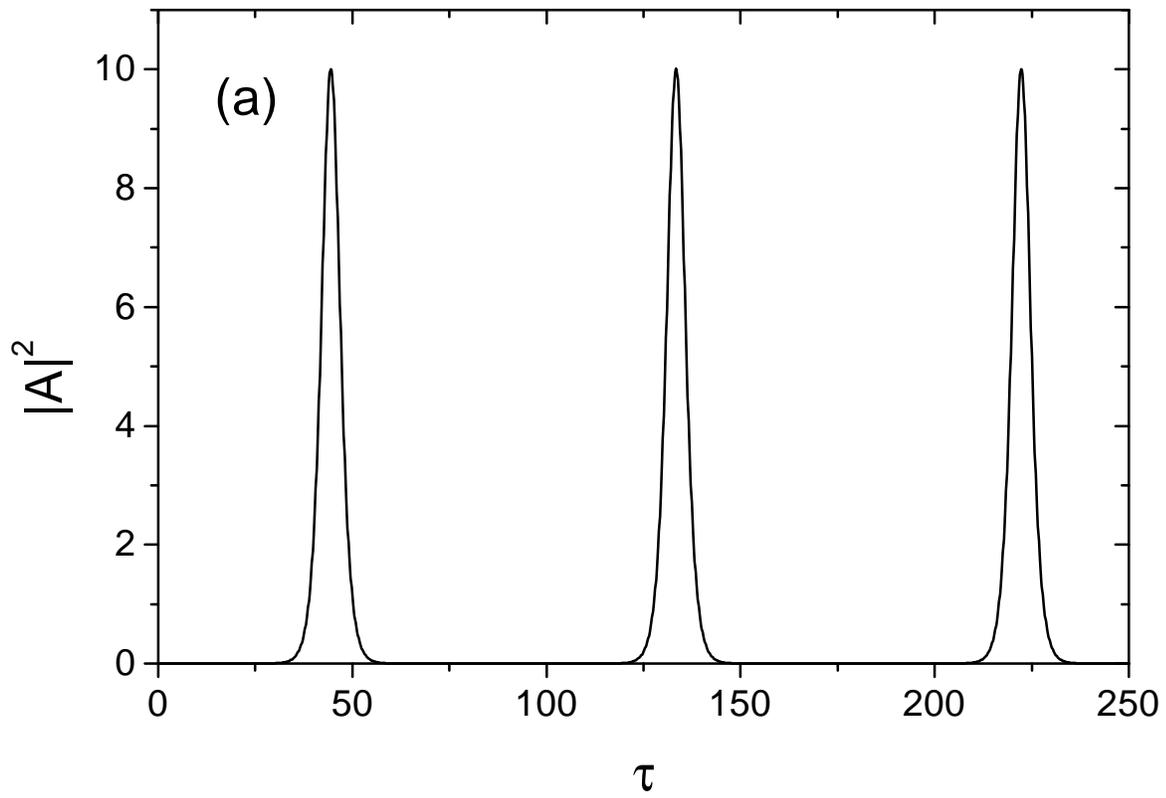
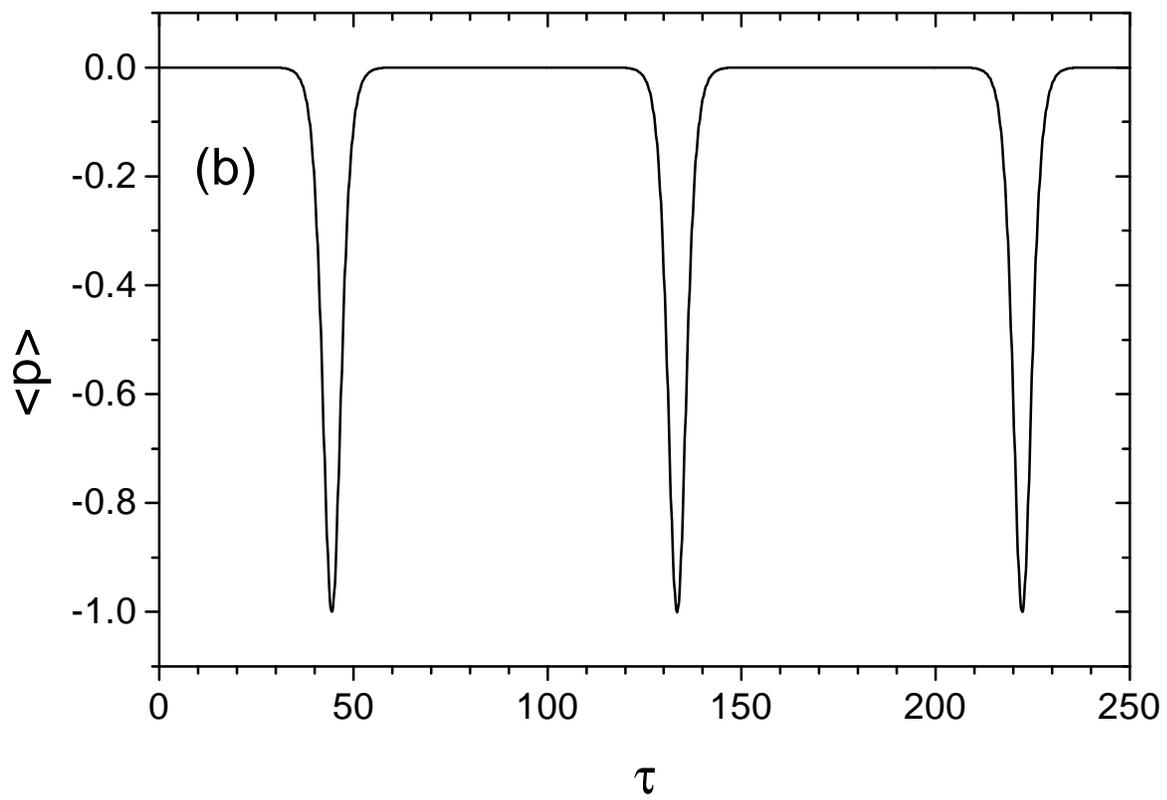
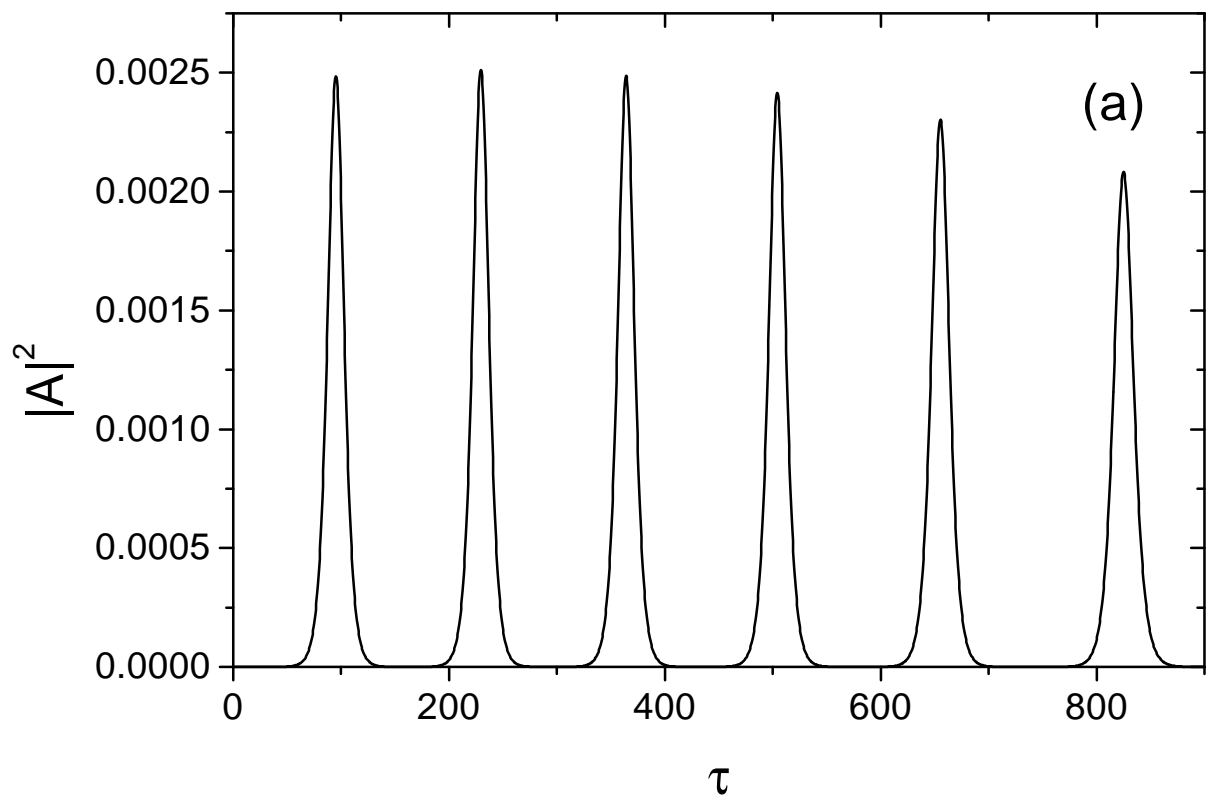
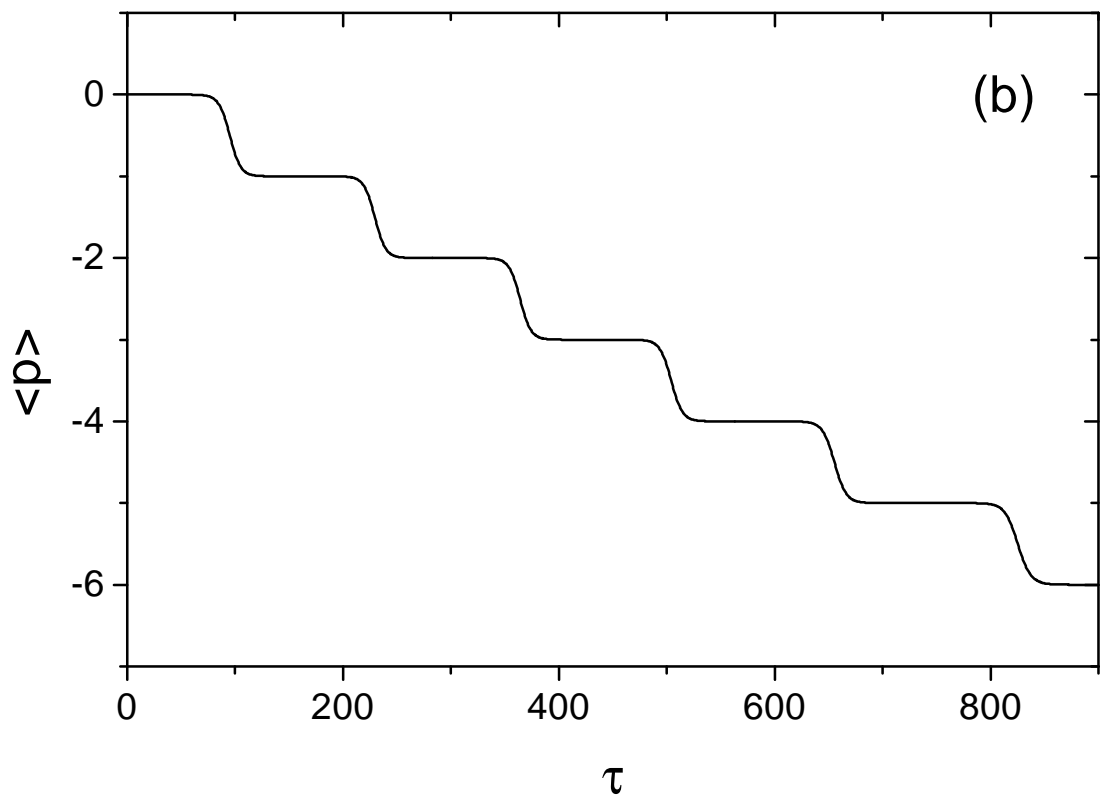
\end{document}